\documentclass[ASNA,twocolumn]{USG} 
\usepackage{anyfontsize} %
\usepackage{siunitx}

\usepackage{colortbl}   
\usepackage{xcolor}     
\usepackage{upgreek}

\usepackage{steinmetz}
\graphicspath{{./images/}}
\specialissue{}

\articletype{RESEARCH ARTICLE}%
\subarticletype{Particle Technology and Fluidization}

\received{XX Month Year}
\revised{XX Month Year}
\accepted{XX Month Year}
\journal{Journal Name}
\volume{0}
\copyyear{Year}
\startpage{1}
\articledoi{10.1002/0000}

\begin{document}

\title{Anomalous Pressure-Enhanced Polarization in Sliding Ferroelectrics}
\transtitle{Anomalous Pressure-Enhanced Polarization in Sliding Ferroelectrics}

\author[1]{Lujin Min}
\author[2]{Sahas Kamat}
\author[3]{Ameer Mustafa}
\author[3]{Nguyen The Duy}
\author[1]{Kyle Nadel}
\author[1]{Junhao Lin}
\author[4]{Kenji Watanabe}
\author[5]{Takashi Taniguchi}
\author[3]{Daniel Bennett}
\author[2]{Brad J. Ramshaw}
\author[1]{Kenji Yasuda}

\authormark{MIN \textsc{et al.}}
\titlemark{Anomalous Pressure-enhanced Polarization in Sliding Ferroelectrics}

\address[1]{\orgdiv{School of Applied and Engineering Physics, }\orgname{Cornell University, }%
\orgaddress{\state{Ithaca, New York 14853, }\country{USA}}}

\address[2]{\orgdiv{Department of Physics, }\orgname{Cornell University, }%
\orgaddress{\state{Ithaca, New York 14853, }\country{USA}}}

\address[3]{\orgdiv{School of Electrical and Electronic Engineering, }\orgname{Nanyang Technological University, }%
\orgaddress{\state{Singapore 639798, }\country{Singapore}}}

\address[4]{\orgdiv{Research Center for Electronic and Optical Materials, }\orgname{National Institute for Materials Science, }%
\orgaddress{\state{Tsukuba, Ibaraki 305-0044, }\country{Japan}}}

\address[5]{\orgdiv{Research Center for Materials Nanoarchitectonics, }\orgname{National Institute for Materials Science, }%
\orgaddress{\state{Tsukuba, Ibaraki 305-0044, }\country{Japan}}}

\corres{Kenji Yasuda (\email{kenji.yasuda@cornell.edu})}

\keywords{sliding ferroelectricity | pressure tuning | van der Waals heterostructures | ferroelectric memory}

\abstract[ABSTRACT]{Hydrostatic pressure commonly suppresses polarization in conventional displacive ferroelectrics by weakening the off-center ionic distortion. Sliding ferroelectrics may provide a contrasting case because their polarization arises from stacking-dependent interlayer charge transfer rather than intra-unit-cell ionic displacement. Here, hydrostatic pressure is shown to enhance, rather than suppress, the polarization of parallel-stacked bilayer boron nitride, a prototypical sliding ferroelectric. A graphene layer placed on top of a ferroelectric boron nitride bilayer is used for sensing the evolution of polarization under pressure. The pressure enhances both the interlayer potential and the ferroelectric polarization, with the latter increasing approximately linearly by $80\%$ up to $1.66~\mathrm{GPa}$. These results establish hydrostatic compression as a tuning parameter for sliding-ferroelectric polarization, with potential implications for ultrathin slidetronic memory devices.}

\maketitle

\section{Introduction}\label{sec:intro}

Ferroelectricity can arise from distinct microscopic mechanisms, leading to different ways in which polarization couples to external perturbations, such as hydrostatic pressure. In conventional displacive ferroelectrics, such as BaTiO$_3$, switchable polarization originates from off-center ionic displacements within a unit cell (Figure~1a). This polar distortion is stabilized by a competition between long-range Coulomb interactions that favor the polar state and short-range interatomic repulsions that favor the centrosymmetric structure \cite{cohen1992origin}. When hydrostatic pressure is applied, it compresses the three-dimensional lattice, enhancing short-range interatomic repulsions more rapidly than the long-range interactions. As a result, the polar distortion is weakened, and the system is driven toward a paraelectric phase under high pressure \cite{samara1971pressure,ishidate1997phase,gourdain1995ferroelectricparaelectric} (Figure~1b).

A fundamentally distinct mechanism has recently emerged in van der Waals layered materials, where polarization arises not from intra-unit-cell ionic displacement but from the interlayer stacking registry \cite{li_binary_2017, woods_charge-polarized_2021,yasuda2021stackingengineered,stern2021interfacial}. For example, when two boron nitride monolayers are assembled with a $0^\circ$ relative orientation, in contrast to $180^\circ$ relative orientation in a bulk crystal, the resulting parallel-stacked bilayer can adopt two energetically degenerate, noncentrosymmetric AB and BA stackings. As shown in Figures~1c and~1d, the reversed vertical alignment of boron and nitrogen atoms in these two stackings induces opposite interlayer charge transfer, producing out-of-plane polarizations that can be switched by a lateral interlayer slide \cite{yasuda2021stackingengineered,stern2021interfacial}. Based on the lattice symmetry, this structure is called rhombohedral boron nitride (rBN). This sliding ferroelectricity, observed in a growing family of two-dimensional systems \cite{fei_ferroelectric_2018, wang2022interfacial,weston2022interfacial,jindal_coupled_2023,sui2023sliding,xue2025emergence}, offers ultrathin, switchable polarization with low switching barriers at room temperature and has opened the prospect of slidetronic non-volatile memory devices \cite{du2024sliding,yasuda2024ultrafast,bian2024developing,yang2024ferroelectric,vareskic2025gatetunable,wang2026resonantenhanced,raab2026sliding}.

Due to its interfacial origin, sliding ferroelectricity may exhibit a pressure response fundamentally different from that of displacive ferroelectrics. In van der Waals materials, the interlayer bonding is much softer than the intralayer covalent bonding, so hydrostatic pressure produces a large reduction of the out-of-plane van der Waals gap while only weakly perturbing the in-plane lattice \cite{rBN14GPa1994, hBN12GPa}. The resulting reduction in the van der Waals gap is expected to strengthen wavefunction overlap between adjacent layers and amplify the stacking-dependent charge redistribution, thereby increasing the out-of-plane polarization (Figures~1c and~1d). This picture is consistent with previous theoretical predictions that reducing the interlayer spacing in bilayer rBN enhances its polarization \cite{li_binary_2017,jiang2022anomalous}. The change in polarization is described as $\Delta P_z \simeq d_{33}\sigma_{zz}$, where $\sigma_{zz}$ is the stress along the out-of-plane direction and $\Delta P_{z}$ is the resulting change in out-of-plane polarization. Here, we neglect the transverse contributions, assuming that the polarization change is governed primarily by the interlayer spacing, which is only weakly affected by in-plane stresses~\cite{bosak2006elasticity}. Applying pressure $p$ corresponds to a negative stress, $\sigma_{zz} = -p$. The out-of-plane piezoelectric coefficient $d_{33}$ is predicted to be negative in bilayer rBN, in contrast to the positive piezoelectricity commonly found in conventional ferroelectrics. Experimentally, however, a quantitative connection between hydrostatic pressure, interlayer charge transfer, and polarization in sliding ferroelectrics has yet to be established.

Here we experimentally probe the pressure dependence of polarization in a sliding ferroelectric by combining graphene-based charge sensing with a hydrostatic-pressure setup. Using bilayer rBN as a model system, we find that hydrostatic pressure enhances the ferroelectric polarization. This pressure-enhanced response is opposite to the pressure-induced suppression commonly observed in conventional displacive ferroelectrics \cite{samara1971pressure,ishidate1997phase,gourdain1995ferroelectricparaelectric} and confirms hydrostatic compression as a direct tuning parameter for interfacial polarization in sliding ferroelectrics. Beyond establishing a unique electromechanical response, our results suggest a route to increasing the readout contrast between binary polarization states, which may be beneficial for future slidetronic devices.

\section{Results}\label{sec2}
\subsection{Device Geometry and Ambient-Pressure Graphene Charge Sensing of Sliding Ferroelectricity}
Figure~2a shows a schematic of the high-pressure measurement setup. The device is mounted on a metal electrical feedthrough inside a Teflon sample tube filled with pressure medium. Insulated copper wires pass through the feedthrough and are sealed with epoxy, allowing electrical access to the device under pressure. Figure~2b shows the device geometry. A monolayer graphene sheet is placed adjacent to a bilayer rBN ferroelectric and encapsulated between multilayer hBN dielectrics. Graphite top and bottom gates provide electrostatic control of the graphene carrier density and the out-of-plane electric field across the ferroelectric bilayer region. 

In this geometry, graphene serves as a charge sensor for the polarization state of the sliding ferroelectric \cite{yasuda2021stackingengineered,wang2022interfacial}. The polarization in bilayer rBN creates a built-in interlayer potential \(\Delta V_{\mathrm{P}}\) between the layers (Figure~2c). When the polarization of the bilayer rBN points toward graphene, an additional gate voltage, \(-\Delta V_{\mathrm{P}}\), is required to compensate the polarization-induced charge and tune graphene to the charge-neutrality point. When the polarization is reversed, the required compensation voltage changes sign to \(+\Delta V_{\mathrm{P}}\). Hence, we can read out the ferroelectric switching as a hysteretic shift of the graphene's charge-neutrality point (CNP), with a forward--backward separation of \(\Delta V_{\mathrm{CNP}} = 2\Delta V_{\mathrm{P}}\). This separation provides a quantitative measure of the built-in interlayer potential of the sliding ferroelectric. Because the width of the graphene resistance peak determines the precision of $\Delta V_{\mathrm{P}}$, we performed the transport measurement at low temperature in a variable temperature insert to minimize the peak broadening.

The bilayer rBN in our device exhibits clear signatures of switchable sliding ferroelectricity at ambient pressure. Figure~2d shows the graphene's longitudinal resistance, \(R_{xx}\), as a function of bottom-gate voltage, \(V_{\mathrm{B}}\). The bottom gate was first swept forward from \(-2~\mathrm{V}\) to \(2~\mathrm{V}\), shown by the red curve, and was then swept backward from \(2~\mathrm{V}\) to \(-2~\mathrm{V}\), shown by the blue curve. A pronounced shift of the graphene charge-neutrality peak is observed between the two sweep directions. As a control, no hysteresis is observed when sweeping the top gate, as shown in Supplementary Section S1, ruling out trivial charge trapping in graphene as the origin. In the forward (backward) scan, the resistance peak appears at \(V_{\mathrm{B}} = 114.8~\mathrm{mV}\) (\(V_{\mathrm{B}} = -102.6~\mathrm{mV}\)). Hence, the forward--backward separation of charge neutrality is \(\Delta V_{\mathrm{CNP}} = 217.4~\mathrm{mV}\), namely, \(\Delta V_{\mathrm{P}} = 108.7~\mathrm{mV}\), consistent with previous reports \cite{woods_charge-polarized_2021, yasuda2021stackingengineered, stern2021interfacial} and our measurement in devices S2 and S3 (Supplementary Section S2). 

In the backward sweep, a weak residual peak remains at about \(V_{\mathrm{B}} = 130~\mathrm{mV}\), around the same place as the charge-neutrality peak position of the forward sweep. This feature suggests that a small fraction of domains associated with the negative-polarization state remains unswitched during the preceding forward sweep. An abrupt change in \(R_{xx}\) appears at \(V_{\mathrm{B}} = 0.92~\mathrm{V}\) in the forward sweep and \(V_{\mathrm{B}} = -0.55~\mathrm{V}\) in the backward sweep, corresponding to the coercive field of ferroelectric switching, as shown in the insets of Figure~2d.

To further resolve the gate dependence of this switching, we mapped \(R_{xx}\) as a function of both bottom- and top-gate voltages. In this dual-gate mapping, we swept \(V_{\mathrm{B}}\) forward from \(-2~\mathrm{V}\) to \(2~\mathrm{V}\) and backward from \(2~\mathrm{V}\) to \(-2~\mathrm{V}\) at each fixed \(V_{\mathrm{T}}\) from \(4~\mathrm{V}\) to \(-4~\mathrm{V}\), as shown in Figures~2e and~2f, respectively. The graphene's charge neutrality appears as high-resistance diagonal lines. Instead of forming a single diagonal line, the lines consist of two parallel branches, exhibiting an abrupt shift at critical \(V_{\mathrm{B}}\). This sudden shift reflects the change in the graphene carrier density caused by the switching of out-of-plane polarization of bilayer rBN. Figure~2g shows the normalized hysteretic component of the resistance, defined as \((R_{xx}^{\mathrm{B}} - R_{xx}^{\mathrm{F}})/(R_{xx}^{\mathrm{B}} + R_{xx}^{\mathrm{F}})\). The switching boundary is nearly \(V_{\mathrm{T}}\)-independent, indicating that the switching is controlled by the electric field applied across the bottom hBN and bilayer rBN. The small \(V_{\mathrm{T}}\) dependence originates from the finite compressibility of graphene, which makes the electric field across the bottom dielectric dependent on graphene's chemical potential \cite{yasuda2021stackingengineered} (see Supplementary Section S3). We also find another switching boundary as we sweep \(V_{\mathrm{B}}\) in the backward direction. The presence of multiple switching boundaries suggests the presence of an intermediate state, where the domain wall is pinned somewhere in the device. Taken together, these features establish the presence of switchable sliding ferroelectricity in bilayer rBN, serving as the ambient-pressure baseline for the pressure-dependent measurements.

\subsection{Pressure-Enhanced Interlayer Potential}

Having established the ferroelectric response of the device at ambient pressure, we next examine how the charge-neutrality-point separation evolves under hydrostatic pressure. Figure~3a shows \(R_{xx}\) as a function of \(V_\mathrm{B}\) at pressures from \(0\) to \(1.66~\mathrm{GPa}\). The ambient-pressure trace was measured at $1.5~\mathrm{K}$, while the finite-pressure traces were measured at $1.7~\mathrm{K}$. This small temperature difference is negligible for the present comparison, as the ferroelectric response has been shown to remain almost temperature-independent over a much wider temperature range~\cite{yasuda2021stackingengineered}. The red and blue traces correspond to forward and backward sweep directions of \(V_\mathrm{B}\), respectively. The separation between the charge-neutrality peaks induced by the two opposite polarization states increases as we increase the pressure from \(0\) to \(1.66~\mathrm{GPa}\). To quantify this trend, we determine the charge-neutrality peak positions by fitting the one-dimensional \(V_{\mathrm{B}}\) scans with Lorentzian line shapes (Supplementary Section S2). The extracted evolution of the interlayer potential \(\Delta V_{\mathrm{P}} = \Delta V_{\mathrm{CNP}}/2\) as a function of pressure is shown in Figure~3b. The interlayer potential increases monotonically from \(109~\mathrm{mV}\) at ambient pressure to \(159~\mathrm{mV}\) at \(1.66~\mathrm{GPa}\), representing a \(46\%\) increase over the measured pressure range. The calculated \(\Delta V_{\mathrm{P}}\) based on the first-principles calculation in Figure~3c exhibits the same trend as the experimental data.

We also find that the domain proportion varies with pressure. At ambient pressure, graphene's resistance shows a single peak in both the backward and forward sweeps, meaning nearly 100\% of the device area switched the polarization direction. Under compression, however, the backward branch contains two resistance peaks, namely, up and down polarization states spatially coexist inside the device. This change of the switching behavior may indicate that the domains become harder to switch under pressure. However, this does not necessarily imply that pressure increases the intrinsic sliding barrier between the two stacking registries. The experimentally observed switching reflects collective domain-wall motion, which can also be influenced by pressure-induced changes in local strain, defect potential, or domain-wall pinning sites. The identification of the pressure evolution of the switching barrier remains future work.

\subsection{Pressure-Enhanced Polarization}

Although the built-in interlayer potential \(\Delta V_{\mathrm{P}}\) is directly quantified from the graphene charge-sensing experiment, the more fundamental quantities of interest for sliding ferroelectricity are interlayer charge transfer and polarization. To extract them quantitatively, we relate \(\Delta V_{\mathrm{P}}\) to the interlayer charge transfer and polarization. Applying Gauss's law to the electrostatic potential profile as shown in Figure~2c, the interlayer charge transfer density $n_\mathrm{CT}$ is described as

\[
n_\mathrm{CT} = \frac{\varepsilon_0 \varepsilon_\mathrm{BN} \Delta V_{\mathrm{P}}}{et}. 
\]

Here, \(t\) is the distance between the two sheets of bilayer rBN, $e$ is the elementary charge, $\varepsilon_0$ is the vacuum permittivity, and $\varepsilon_\mathrm{BN}$ is the relative out-of-plane dielectric constant of boron nitride, which we assume to be the same for multilayer hBN and bilayer rBN. As the capacitance between the graphene and bottom graphite is written as 

\[
C_\mathrm{B} = \frac{\varepsilon_0 \varepsilon_\mathrm{BN}}{d_\mathrm{B}} ,
\]

where $d_\mathrm{B}$ is the distance between the graphene and bottom graphite, the interlayer charge-transfer density can be described as 

\[
n_\mathrm{CT} = \frac{C_\mathrm{B} \Delta V_{\mathrm{P}}}{e}\frac{d_\mathrm{B}}{t}.
\]

This is also related to the polarization $P$, expressed in the conventional three-dimensional polarization unit of $\mu\mathrm{C\,cm^{-2}}$, by a simple relation

\[
P = en_\mathrm{CT} = C_\mathrm{B} \Delta V_{\mathrm{P}}\frac{d_\mathrm{B}}{t}.
\]

Previous high-pressure X-ray diffraction measurements show that hBN and rBN exhibit closely comparable \(c\)-axis compression over the pressure range relevant here \cite{rBN14GPa1994, hBN12GPa}, such that the geometric factor \(d_\mathrm{B}/t\) can be treated as pressure-independent. Thus, the interlayer charge-transfer density \(n_{\mathrm{CT}}\) and polarization $P$ can be estimated from the measured \(\Delta V_{\mathrm{P}}\) and the pressure dependence of the bottom capacitance \(C_\mathrm{B}\). 

We performed Landau-fan measurements to extract the pressure evolution of the bottom capacitance \(C_\mathrm{B}\). Because the ferroelectric switching produces multiple charge-neutrality features in the backward sweep at high pressure (Figure~3a), we swept $V_\mathrm{B}$ in the forward direction to obtain a Landau fan emanating from a single \(V_{\mathrm{B}}\) point. Figure~4a shows the forward-sweep Landau fan of graphene at \(1.66~\mathrm{GPa}\) and \(1.7~\mathrm{K}\). Well-defined Landau-level trajectories emerge from the charge-neutrality region. Their noticeable bending away from a simple linear relation in $V_\mathrm{B}$ reflects the finite compressibility of graphene, i.e., the quantum-capacitance correction.

To visualize the pressure-induced change in capacitance, we plot one-dimensional line cuts at $B=6.02~\mathrm{T}$ for four different finite pressures, together with the ambient-pressure data measured at $B=6~\mathrm{T}$ in Figure~4b. The traces are vertically offset for clarity. The peaks in the Shubnikov-de Haas oscillations shift systematically toward smaller \(|V_\mathrm{B}-V_{\mathrm{CNP}}|\) with increasing pressure, indicating that a smaller gate voltage is required to reach the same carrier density. Namely, the pressure increases the bottom-gate capacitance \(C_\mathrm{B}\).

By fitting the Landau-level positions at each pressure, we extract the pressure dependence of \(C_\mathrm{B}\), as shown in Figure~4c. We find that the capacitance increases at an average rate of approximately \(14\%\) per GPa, comparable to the previously reported value of approximately \(9\%\) per GPa in hBN-based graphene devices, which was attributed to the combined effects of dielectric-constant enhancement and thickness reduction under pressure \cite{yankowitz2018dynamic}.

The geometric factor \(d_\mathrm{B}/t\) is determined from the topography obtained with atomic force microscopy. The bottom multilayer hBN thickness is approximately \(4.0~\mathrm{nm}\) (see Supplementary Section S4), corresponding to \(12\) hBN layers. Including bilayer rBN, it gives \(d_\mathrm{B}/t\) of 14. Combining this geometric factor with the extracted \(C_\mathrm{B}\) values and the measured \(\Delta V_{\mathrm{P}}\), we obtain the interlayer charge-transfer density \(n_{\mathrm{CT}}\) and polarization $P$, plotted in Figure~4d. The interlayer charge-transfer density \(n_{\mathrm{CT}}\) increases from \(4.46\times10^{12}~\mathrm{cm^{-2}}\) at \(0~\mathrm{GPa}\) to \(8.03\times10^{12}~\mathrm{cm^{-2}}\) at \(1.66~\mathrm{GPa}\), representing an \(80\%\) increase over the measured pressure range, or an average increase of \(48\%\) per GPa. Correspondingly, the polarization increases from \(0.71~\mu\mathrm{C\,cm^{-2}}\) to \(1.29~\mu\mathrm{C\,cm^{-2}}\). Meanwhile, based on previous high-pressure X-ray diffraction measurements of rBN, $1.66~\mathrm{GPa}$ corresponds to an approximately $3.9\%$ reduction in the rBN interlayer spacing~\cite{rBN14GPa1994}. Since both the polarization and the interlayer spacing~\cite{rBN14GPa1994} vary approximately linearly with pressure over the measurement range, their relationship is also nearly linear, with a $1\%$ reduction in spacing corresponding to a $\sim21\%$ increase in polarization.

Under the conventional sign convention in which tensile stress is positive, the applied pressure corresponds to a compressive stress, \(\sigma_{zz}=-p\). Therefore, the increase in polarization under pressure yields a negative out-of-plane piezoelectric coefficient,
\[
d_{33}
\simeq
\frac{\Delta P_{z}}{\Delta \sigma_{zz}}
=
-\frac{\Delta P_{z}}{\Delta p}
\approx
-3.5~\mathrm{pC\,N^{-1}}.
\]

\section{Discussion and Conclusion}\label{sec3}
Negative longitudinal piezoelectricity is extremely rare among ferroelectrics \cite{katsouras_negative_2016,liu_origin_2017,qi_widespread_2021}, but was recently observed in a layered ferroelectric, CuInP$_2$S$_6$ \cite{you_origin_2019, yao_anomalous_2023}. In CuInP$_2$S$_6$, the experimentally observed pressure-enhanced polarization is tied to an ionic mechanism: compression of the van der Waals gap drives the redistribution of Cu ions \cite{you_origin_2019, yao_anomalous_2023}. The negative piezoelectric response of bilayer rBN originates from a qualitatively distinct mechanism. Here, the polarization originates from interlayer charge transfer, without polar ionic displacements within the layers. Because this charge transfer is governed by interlayer orbital hybridization, it is enhanced in response to the compression of the van der Waals gap. The negative sign of $d_{33}$ is therefore not governed by ionic displacement, but is a consequence of the electronic, interlayer origin of polarization in sliding ferroelectrics. This qualitative distinction also leads to different pressure dependences. In CuInP$_2$S$_6$, the polarization is enhanced below 0.26 GPa, remains nearly constant over an intermediate pressure range, and is eventually suppressed at higher pressure. By contrast, the polarization in rBN increases monotonically over the pressure range studied here, due to the gradual decrease in the interlayer distance. Because of the interlayer charge-transfer nature, we anticipate that pressure-enhanced polarization is universal among other sliding ferroelectrics beyond rBN \cite{ding_phase_2021,xu_how_2024}.

The pressure-enhanced polarization observed here may benefit sliding-ferroelectric memory devices. In both ferroelectric field-effect transistors \cite{yasuda2024ultrafast,bian2024developing} and ferroelectric tunneling devices \cite{vareskic2025gatetunable,wang2026resonantenhanced,raab2026sliding}, a larger polarization is expected to increase the readout contrast between the two polarization states. Pressure-enhanced polarization may therefore improve the readout margin and facilitate reliable discrimination between the up- and down-polarized states. Although hydrostatic pressure is used as the experimental tuning parameter in this study, the underlying microscopic effect is a reduction of the out-of-plane van der Waals gap. This suggests that the same polarization enhancement could be achieved through any strain-engineering approach that compresses the interlayer spacing. Indeed, SiN$_x$ stressor layers have recently been shown to generate local cross-plane stresses up to hundreds of MPa in scaled two-dimensional devices \cite{jaikissoon2024cmoscompatible}, suggesting the possibility of translating pressure-enhanced sliding ferroelectricity found here into practical slidetronic devices.

\section{Methods}\label{sec4}

\subsection{Device Fabrication}

The heterostructure was fabricated using a dry-transfer method. Monolayer graphene, graphite, and multilayer hBN flakes were exfoliated onto SiO\(_2\)/Si substrates with a \(285~\mathrm{nm}\) SiO\(_2\) layer. Monolayer hBN flakes were used to prepare the sliding-ferroelectric bilayer BN and were exfoliated separately on SiO\(_2\)/Si substrates with a \(90~\mathrm{nm}\) SiO\(_2\) layer for ease of identification. Flakes were sequentially picked up using a poly(bisphenol A carbonate)-covered polydimethylsiloxane stamp (PC/PDMS) on a motorized transfer stage.

The bottom-gate stack, consisting of multilayer hBN and graphite, was first assembled, released onto a \(2~\mathrm{mm}~\times~2~\mathrm{mm}\) SiO\(_2\)/intrinsic Si substrate, and cleaned by contact-mode atomic force microscopy. The top stack was then assembled by sequentially picking up graphite, multilayer hBN, monolayer graphene, and a parallel-stacked bilayer BN prepared by the tear-and-stack method \cite{kim2016van,cao2016superlatticeinduced}. For the bilayer BN, one half of a monolayer BN flake was picked up first, followed by the remaining half without rotating the stage. The completed upper stack was subsequently released onto the bottom-gate stack. The device was patterned by electron-beam lithography and reactive ion etching. One-dimensional edge contacts were made by reactive ion etching and evaporating Cr/Au (\(10/40~\mathrm{nm}\)) at a \(15^\circ\) tilt angle.

\subsection{High-Pressure Measurement Preparation}

High-pressure measurements were performed using a CTF-HHPC60 piston-cylinder pressure cell made by C\&T Factory Co., Ltd. Insulated copper wires were inserted through a clean metal feedthrough and fixed in place using Stycast 2850FT epoxy with 23LV catalyst. The device, fabricated on a \(2~\mathrm{mm}\times 2~\mathrm{mm}\) SiO\(_2\)/intrinsic Si substrate, was then mounted on top of the cured epoxy. Electrical connections between the device and the copper wires were made using Pt wires bonded with CircuitWorks CW2400 room-temperature silver epoxy. The device was encapsulated in Daphne 7373 oil \cite{murata1997pt} by filling a Teflon sample tube with the pressure medium and fitting it over the device onto the feedthrough. The feedthrough assembly was inserted into the inner bore of the piston-cylinder cell. The desired load was applied using a hydraulic press, and the locking nut was tightened under load to maintain the pressure after the press was released. The pressure cell was mounted on a modified probe for the Oxford Instruments TeslatronPT cryostat.

The measurement device was pressurized at room temperature according to the hydraulic-press reading. The corresponding room-temperature pressure was determined using the previously established manganin calibration of the same pressure cell \cite{kamatthermodynamic2026}. The low-temperature pressure was estimated as \(p_{\mathrm{LT}}=p_{\mathrm{RT}}-0.15~\mathrm{GPa}\) to account for the pressure loss upon cooling with Daphne 7373 \cite{murata1997pt}.

\subsection{Low-Temperature Transport Measurement}

Low-temperature transport measurements were performed in a variable-temperature insert in a \(^4\)He cryostat, Oxford Instruments TeslatronPT. Longitudinal and Hall resistances, \(R_{xx}\) and \(R_{yx}\), were measured using standard four-probe lock-in techniques on Hall-bar devices with SRS SR860 lock-in amplifiers at \(21.777~\mathrm{Hz}\). An excitation current of \(100~\mathrm{nA}\) was used for high-pressure measurements, whereas \(1~\mathrm{nA}\) was used for ambient-pressure measurements. For ambient-pressure measurements, a Basel Precision Instruments SP983c-IF current preamplifier and Basel Precision Instruments SP1004 voltage preamplifiers were used to reduce noise. The top- and bottom-gate voltages, \(V_\mathrm{T}\) and \(V_\mathrm{B}\), were supplied using Keithley 2450 sourcemeters.

\subsection{DFT Calculation}

First-principles density functional theory (DFT) calculations were performed with the \texttt{SIESTA} \cite{solerSIESTAMethodInitio2002} code, using PSML norm-conserving pseudopotentials \cite{garciaPsmlFormatLibrary2018} obtained from \texttt{Pseudo-Dojo} \cite{pseudodojo}. 
\texttt{SIESTA} employs a basis set of numerical atomic orbitals (NAOs) \cite{junqueraNumericalAtomicOrbitals2001}, and an optimized `DZPF' basis set for hBN from Ref.~\cite{bennettAccurateEfficientLocalized2025} was used, which includes two radial basis functions per valence electron, as well as single $3d$ and $4f$ polarized orbitals for angular flexibility.
A Monkhorst-Pack $k$-point grid of $15 \times 15 \times 1$ was used in all calculations, and a fine real-space grid was used, determined using a mesh cutoff of 1200 Ry.
We used the PBE exchange-correlation functional \cite{pbe} in all calculations and included a DFT-D3 dispersion correction \cite{grimme2010consistent} to treat long-range interactions between layers. 
Calculations were converged until the relative changes in the Hamiltonian and density matrix were less than $10^{-7}$ and $10^{-6}$, respectively.

The monolayer was fully relaxed first to obtain the optimized cell parameters, using a force tolerance of $0.1~\mathrm{meV/\mathring{A}}$. 
The out-of-plane lattice constant was fixed at 40 Å to separate the bilayer from its periodic images, and a dipole correction was employed in the vacuum region to prevent spurious interactions.
A series of calculations was performed for bilayer rBN, varying the interlayer separation $d$, and the free energy was fit to the Birch-Murnaghan equation of state:
\begin{align*}
 E(d) = E_0 + \frac{9}{16}\,V_f B\Bigg\{
  &\left[\left(\frac{d_0}{d}\right)^{2/3}-1\right]^3 B' \\
  &+ \left[\left(\frac{d_0}{d}\right)^{2/3}-1\right]^2
     \left[6 - 4\left(\tfrac{d_0}{d}\right)^{2/3}\right] \Bigg\},
\end{align*}
from which the equilibrium layer separation $d_0$, bulk modulus $B$ and its derivative $B'$ were calculated.
The pressure $p(d)$ is given by \cite{zhuWavefunctionTexturesTwisted2026}:
\begin{equation}
  p \left( d \right) = -B \frac{\left( d - d_{0} \right)}{d_{0}},
\end{equation}
which enables the tuning of pressure by changing the interlayer separation.
The out-of-plane dipole moment and electrostatic potential drop across the bilayer were then obtained as a function of pressure.
The differential charge density was obtained by calculating the charge density of the bilayer and subtracting the charge densities of the individual monolayers in isolation, using the \texttt{c2x} utility \cite{rutterC2xToolVisualisation2018}.

\bmsubsection*{Author Contributions}

K.Y. conceived and supervised the project. L.M., K.N., and J.L. stacked and fabricated the 2D device. L.M., S.K., and B.J.R. performed the high-pressure measurements. A.M., N.T.D., and D.B. performed DFT calculations. K.W. and T.T. grew the bulk hBN crystal. L.M. and K.Y. wrote the manuscript with input from all other authors.

\bmsubsection*{Acknowledgments}

We thank Matthew Yankowitz for helpful advice and discussions regarding high-pressure experiments. This work was supported by the Office of Naval Research (ONR) under Award No. N00014-26-1-2259.
This work was performed in part at the Cornell NanoScale Facility, a member of the National Nanotechnology Coordinated Infrastructure (NNCI), which is supported by the National Science Foundation (Grant NNCI-2025233).
This work made use of the Cornell Center for Materials Research shared instrumentation facility.
K.W. and T.T. acknowledge support from JSPS KAKENHI (Grant Nos. 21H05233 and 23H02052), CREST (JPMJCR24A5), JST, and the World Premier International Research Center Initiative (WPI), MEXT, Japan.
D.B., A.M.~and N.T.D.~acknowledge support from the NTU Startup Grant (Award No. 025661-00003).
B.J.R. and S.K. acknowledge support from the U.S. Department of Energy, Office of Basic Energy Sciences, under Award No. DE-SC-0026003 (high pressure measurements). 

\bmsubsection*{Conflicts of Interest}

The authors declare no conflicts of interest.

\bmsubsection*{Data Availability Statement}

The data used in this manuscript are openly available in Zenodo at \href{https://doi.org/10.5281/zenodo.22923946}{10.5281/zenodo.22923946}.

\bibliography{reference_new}

\providecommand{\noopsort}[1]{}
\begin{thebibliography}{47}
\providecommand{\natexlab}[1]{#1}
\providecommand{\url}[1]{\texttt{#1}}
\expandafter\ifx\csname urlstyle\endcsname\relax
  \providecommand{\doi}[1]{doi: #1}\else
  \providecommand{\doi}{doi: \begingroup \urlstyle{rm}\Url}\fi

\bibitem[\protect\citeauthoryear{Cohen}{Cohen}{}]{cohen1992origin}
Ronald~E. Cohen.
\newblock
\newblock ``Origin of Ferroelectricity in Perovskite Oxides.''  {\it Nature\/}~358, no. 6382 (1992): 136--138.

\bibitem[\protect\citeauthoryear{Samara}{Samara}{}]{samara1971pressure}
G.~A. Samara.
\newblock
\newblock ``Pressure and Temperature Dependence of the Dielectric Properties and Phase Transitions of the Ferroelectric Perovskites: {{PbTiO3}} and {{BaTiO3}}.''  {\it Ferroelectrics\/}~2, no. 1 (1971): 277--289.

\bibitem[\protect\citeauthoryear{Ishidate, Abe, Takahashi, and M{\^o}ri}{Ishidate et~al.}{}]{ishidate1997phase}
T.~Ishidate, S.~Abe, H.~Takahashi, and N.~M{\^o}ri.
\newblock
\newblock ``Phase {{Diagram}} of {{BaTiO}}{$_{3}$}.''  {\it Physical Review Letters\/}~78, no. 12 (1997): 2397--2400.

\bibitem[\protect\citeauthoryear{Gourdain, Moya, Chervin, Canny, and Pruzan}{Gourdain et~al.}{}]{gourdain1995ferroelectricparaelectric}
D.~Gourdain, E.~Moya, J.~C. Chervin, B.~Canny, and {\relax Ph}.~Pruzan.
\newblock
\newblock ``Ferroelectric-Paraelectric Phase Transition in {{KNbO}}{$_3$} at High Pressure.''  {\it Physical Review B\/}~52, no. 5 (1995): 3108--3112.

\bibitem[\protect\citeauthoryear{Li and Wu}{Li and Wu}{}]{li_binary_2017}
Lei Li and Menghao Wu.
\newblock
\newblock ``Binary {Compound} {Bilayer} and {Multilayer} with {Vertical} {Polarizations}: {Two}-{Dimensional} {Ferroelectrics}, {Multiferroics}, and {Nanogenerators}.''  {\it ACS Nano\/}~11, no. 6 (2017): 6382--6388.

\bibitem[\protect\citeauthoryear{Woods, Ares, Nevison-Andrews, Holwill, Fabregas, Guinea, Geim, Novoselov, Walet, and Fumagalli}{Woods et~al.}{}]{woods_charge-polarized_2021}
C.~R. Woods, P.~Ares, H.~Nevison-Andrews, M.~J. Holwill, R.~Fabregas, F.~Guinea, A.~K. Geim, K.~S. Novoselov, N.~R. Walet, and L.~Fumagalli.
\newblock
\newblock ``Charge-polarized interfacial superlattices in marginally twisted hexagonal boron nitride.''  {\it Nature Communications\/}~12, no. 1 (2021): 347.

\bibitem[\protect\citeauthoryear{Yasuda, Wang, Watanabe, Taniguchi, and {Jarillo-Herrero}}{Yasuda et~al.}{}]{yasuda2021stackingengineered}
Kenji Yasuda, Xirui Wang, Kenji Watanabe, Takashi Taniguchi, and Pablo {Jarillo-Herrero}.
\newblock
\newblock ``Stacking-Engineered Ferroelectricity in Bilayer Boron Nitride.''  {\it Science\/}~372, no. 6549 (2021): 1458--1462.

\bibitem[\protect\citeauthoryear{Stern, Waschitz, Cao, Nevo, Watanabe, Taniguchi, Sela, Urbakh, Hod, and Shalom}{Stern et~al.}{}]{stern2021interfacial}
M.~Vizner Stern, Y.~Waschitz, W.~Cao, I.~Nevo, K.~Watanabe, T.~Taniguchi, E.~Sela, M.~Urbakh, O.~Hod, and M.~Ben Shalom.
\newblock
\newblock ``Interfacial Ferroelectricity by van Der {{Waals}} Sliding.''  {\it Science\/}~372, no. 6549 (2021): 1462--1466.

\bibitem[\protect\citeauthoryear{Fei, Zhao, Palomaki, Sun, Miller, Zhao, Yan, Xu, and Cobden}{Fei et~al.}{}]{fei_ferroelectric_2018}
Zaiyao Fei, Wenjin Zhao, Tauno~A. Palomaki, Bosong Sun, Moira~K. Miller, Zhiying Zhao, Jiaqiang Yan, Xiaodong Xu, and David~H. Cobden.
\newblock
\newblock ``Ferroelectric switching of a two-dimensional metal.''  {\it Nature\/}~560, no. 7718 (2018): 336--339.

\bibitem[\protect\citeauthoryear{Wang, Yasuda, Zhang, Liu, Watanabe, Taniguchi, Hone, Fu, and {Jarillo-Herrero}}{Wang et~al.}{}]{wang2022interfacial}
Xirui Wang, Kenji Yasuda, Yang Zhang, Song Liu, Kenji Watanabe, Takashi Taniguchi, James Hone, Liang Fu, and Pablo {Jarillo-Herrero}.
\newblock
\newblock ``Interfacial Ferroelectricity in Rhombohedral-Stacked Bilayer Transition Metal Dichalcogenides.''  {\it Nature Nanotechnology\/}~17, no. 4 (2022): 367--371.

\bibitem[\protect\citeauthoryear{Weston, Castanon, Enaldiev, Ferreira, Bhattacharjee, Xu, {Corte-Le{\'o}n}, Wu, Clark, Summerfield, Hashimoto, Gao, Wang, Hamer, Read, Fumagalli, Kretinin, Haigh, Kazakova, Geim, Fal'ko, and Gorbachev}{Weston et~al.}{}]{weston2022interfacial}
Astrid Weston, Eli~G. Castanon, Vladimir Enaldiev, F{\'a}bio Ferreira, Shubhadeep Bhattacharjee, Shuigang Xu, H{\'e}ctor {Corte-Le{\'o}n}, Zefei Wu, Nicholas Clark, Alex Summerfield, Teruo Hashimoto, Yunze Gao, Wendong Wang, Matthew Hamer, Harriet Read, Laura Fumagalli, Andrey~V. Kretinin, Sarah~J. Haigh, Olga Kazakova, A.~K. Geim, Vladimir~I. Fal'ko, and Roman Gorbachev.
\newblock
\newblock ``Interfacial Ferroelectricity in Marginally Twisted {{2D}} Semiconductors.''  {\it Nature Nanotechnology\/}~17, no. 4 (2022): 390--395.

\bibitem[\protect\citeauthoryear{Jindal, Saha, Li, Taniguchi, Watanabe, Hone, Birol, Fernandes, Dean, Pasupathy, and Rhodes}{Jindal et~al.}{}]{jindal_coupled_2023}
Apoorv Jindal, Amartyajyoti Saha, Zizhong Li, Takashi Taniguchi, Kenji Watanabe, James~C. Hone, Turan Birol, Rafael~M. Fernandes, Cory~R. Dean, Abhay~N. Pasupathy, and Daniel~A. Rhodes.
\newblock
\newblock ``Coupled ferroelectricity and superconductivity in bilayer {Td}-{MoTe2}.''  {\it Nature\/}~613, no. 7942 (2023): 48--52.

\bibitem[\protect\citeauthoryear{Sui, Jin, Zhang, Qi, Wu, Huang, Yue, and Chu}{Sui et~al.}{}]{sui2023sliding}
Fengrui Sui, Min Jin, Yuanyuan Zhang, Ruijuan Qi, Yu-Ning Wu, Rong Huang, Fangyu Yue, and Junhao Chu.
\newblock
\newblock ``Sliding Ferroelectricity in van Der {{Waals}} Layered {$\gamma$}-{{InSe}} Semiconductor.''  {\it Nature Communications\/}~14, no. 1 (2023): 36.

\bibitem[\protect\citeauthoryear{Xue, Wang, Ci, Guo, Qu, Zeng, Liu, He, Cheng, and Xu}{Xue et~al.}{}]{xue2025emergence}
Wuhong Xue, Peng Wang, Wenjuan Ci, Ying Guo, Jingyuan Qu, Zeting Zeng, Tianqi Liu, Ri~He, Shaobo Cheng, and Xiaohong Xu.
\newblock
\newblock ``Emergence of Sliding Ferroelectricity in Naturally Parallel-Stacked Multilayer {{ReSe2}} Semiconductor.''  {\it Nature Communications\/}~16, no. 1 (2025): 6313.

\bibitem[\protect\citeauthoryear{Du, Yang, Gao, Dong, Xu, Watanabe, Taniguchi, Zhao, Zheng, Zhou, and Zheng}{Du et~al.}{}]{du2024sliding}
Shuang Du, Wenqi Yang, Huiying Gao, Weikang Dong, Boyu Xu, Kenji Watanabe, Takashi Taniguchi, Jing Zhao, Fawei Zheng, Jiadong Zhou, and Shoujun Zheng.
\newblock
\newblock ``Sliding {{Memristor}} in {{Parallel-Stacked Hexagonal Boron Nitride}}.''  {\it Advanced Materials\/}~36, no. 35 (2024): 2404177.

\bibitem[\protect\citeauthoryear{Yasuda, {Zalys-Geller}, Wang, Bennett, Cheema, Watanabe, Taniguchi, Kaxiras, {Jarillo-Herrero}, and Ashoori}{Yasuda et~al.}{}]{yasuda2024ultrafast}
Kenji Yasuda, Evan {Zalys-Geller}, Xirui Wang, Daniel Bennett, Suraj~S. Cheema, Kenji Watanabe, Takashi Taniguchi, Efthimios Kaxiras, Pablo {Jarillo-Herrero}, and Raymond Ashoori.
\newblock
\newblock ``Ultrafast High-Endurance Memory Based on Sliding Ferroelectrics.''  {\it Science\/}~385, no. 6704 (2024): 53--56.

\bibitem[\protect\citeauthoryear{Bian, He, Pan, Li, Cao, Meng, Chen, Liu, Zhong, Li, and Liu}{Bian et~al.}{}]{bian2024developing}
Renji Bian, Ri~He, Er~Pan, Zefen Li, Guiming Cao, Peng Meng, Jiangang Chen, Qing Liu, Zhicheng Zhong, Wenwu Li, and Fucai Liu.
\newblock
\newblock ``Developing Fatigue-Resistant Ferroelectrics Using Interlayer Sliding Switching.''  {\it Science\/}~385, no. 6704 (2024): 57--62.

\bibitem[\protect\citeauthoryear{Yang, Liang, Hu, Chen, Ho, Chang, Yang, Lo, Kuo, Chen, Lin, Simbulan, Luo, Chang, Kuo, Ku, Chen, Huang, Chang, Chiang, Lu, Lee, Li, Wu, Chen, Lin, and Lan}{Yang et~al.}{}]{yang2024ferroelectric}
Tilo~H. Yang, Bor-Wei Liang, Hsiang-Chi Hu, Fu-Xiang Chen, Sheng-Zhu Ho, Wen-Hao Chang, Liu Yang, Han-Chieh Lo, Tzu-Hao Kuo, Jyun-Hong Chen, Po-Yen Lin, Kristan~Bryan Simbulan, Zhao-Feng Luo, Alice~Chinghsuan Chang, Yi-Hao Kuo, Yu-Seng Ku, Yi-Cheng Chen, You-Jia Huang, Yu-Chen Chang, Yu-Fan Chiang, Ting-Hua Lu, Min-Hung Lee, Kai-Shin Li, Menghao Wu, Yi-Chun Chen, Chun-Liang Lin, and Yann-Wen Lan.
\newblock
\newblock ``Ferroelectric Transistors Based on Shear-Transformation-Mediated Rhombohedral-Stacked Molybdenum Disulfide.''  {\it Nature Electronics\/}~7, no. 1 (2024): 29--38.

\bibitem[\protect\citeauthoryear{Vareskic, Kennedy, Taniguchi, Watanabe, Yasuda, and Ralph}{Vareskic et~al.}{}]{vareskic2025gatetunable}
Bozo Vareskic, Finn~G. Kennedy, Takashi Taniguchi, Kenji Watanabe, Kenji Yasuda, and Daniel~C. Ralph.
\newblock
\newblock ``Gate-{{Tunable Electroresistance}} in a {{Sliding Ferroelectric Tunnel Junction}}.''  {\it Nano Letters\/}~25, no. 51 (2025): 17540--17546.

\bibitem[\protect\citeauthoryear{Wang, Chen, Pan, Wang, Li, Yang, Zhou, Xie, Liu, Luo, Chu, Li, and Liu}{Wang et~al.}{}]{wang2026resonantenhanced}
Ruixue Wang, Jiangang Chen, Er~Pan, Wunan Wang, Zefen Li, Fan Yang, Hongmiao Zhou, Zhaoren Xie, Qing Liu, Xiao Luo, Junhao Chu, Wenwu Li, and Fucai Liu.
\newblock ``Resonant-Enhanced Tunneling Electroresistance in Sliding Ferroelectric Tunnel Junctions.''  .
\newblock {\it arXiv preprint arXiv:2603.28482\/}.

\bibitem[\protect\citeauthoryear{Raab, Yadav, Bloch, Yeo, Maoz, Plutnarova, Sofer, Watanabe, Taniguchi, and Shalom}{Raab et~al.}{}]{raab2026sliding}
Noam Raab, Renu Yadav, Yakov Bloch, Youngki Yeo, Chen Maoz, Iva Plutnarova, Zdenek Sofer, Kenji Watanabe, Takashi Taniguchi, and Moshe~Ben Shalom.
\newblock ``A Sliding Ferroelectric Resonant Tunnel Junction.''  .
\newblock {\it arXiv preprint arXiv:2603.00817\/}.

\bibitem[\protect\citeauthoryear{Solozhenko, Will, Hüpen, and Elf}{Solozhenko et~al.}{}]{rBN14GPa1994}
V.L. Solozhenko, G.~Will, H.~Hüpen, and F.~Elf.
\newblock
\newblock ``Isothermal compression of rhombohedral boron nitride up to 14 GPa.''  {\it Solid State Communications\/}~90, no. 1 (1994): 65--67.

\bibitem[\protect\citeauthoryear{Solozhenko, Will, and Elf}{Solozhenko et~al.}{}]{hBN12GPa}
V.L Solozhenko, G~Will, and F~Elf.
\newblock
\newblock ``Isothermal compression of hexagonal graphite-like boron nitride up to 12 GPa.''  {\it Solid State Communications\/}~96, no. 1 (1995): 1--3.

\bibitem[\protect\citeauthoryear{Jiang, Liu, Ma, Yu, Hu, Li, Burton, Liu, Chen, Guo, Kong, Bellaiche, and Ren}{Jiang et~al.}{}]{jiang2022anomalous}
Wen Jiang, Chang Liu, Xiaonan Ma, Xing Yu, Shunbo Hu, Xi~Li, Lee~A. Burton, Yu~Liu, Yangyang Chen, Pan Guo, Xiangyang Kong, Laurent Bellaiche, and Wei Ren.
\newblock
\newblock ``Anomalous Ferroelectricity and Double-Negative Effects in Bilayer Hexagonal Boron Nitride.''  {\it Physical Review B\/}~106, no. 5 (2022): 054104.

\bibitem[\protect\citeauthoryear{Bosak, Serrano, Krisch, Watanabe, Taniguchi, and Kanda}{Bosak et~al.}{}]{bosak2006elasticity}
Alexey Bosak, Jorge Serrano, Michael Krisch, Kenji Watanabe, Takashi Taniguchi, and Hisao Kanda.
\newblock
\newblock ``Elasticity of hexagonal boron nitride: Inelastic x-ray scattering measurements.''  {\it Phys. Rev. B\/}73 (2006): 041402(R).

\bibitem[\protect\citeauthoryear{Yankowitz, Jung, Laksono, Leconte, Chittari, Watanabe, Taniguchi, Adam, Graf, and Dean}{Yankowitz et~al.}{}]{yankowitz2018dynamic}
Matthew Yankowitz, Jeil Jung, Evan Laksono, Nicolas Leconte, Bheema~L. Chittari, K.~Watanabe, T.~Taniguchi, Shaffique Adam, David Graf, and Cory~R. Dean.
\newblock
\newblock ``Dynamic Band-Structure Tuning of Graphene Moir\'e Superlattices with Pressure.''  {\it Nature\/}~557, no. 7705 (2018): 404--408.

\bibitem[\protect\citeauthoryear{Katsouras, Asadi, Li, van Driel, Kjær, Zhao, Lenz, Gu, Blom, Damjanovic, Nielsen, and de~Leeuw}{Katsouras et~al.}{}]{katsouras_negative_2016}
Ilias Katsouras, Kamal Asadi, Mengyuan Li, Tim~B. van Driel, Kasper~S. Kjær, Dong Zhao, Thomas Lenz, Yun Gu, Paul W.~M. Blom, Dragan Damjanovic, Martin~M. Nielsen, and Dago~M. de~Leeuw.
\newblock
\newblock ``The negative piezoelectric effect of the ferroelectric polymer poly(vinylidene fluoride).''  {\it Nature Materials\/}~15, no. 1 (2016): 78--84.

\bibitem[\protect\citeauthoryear{Liu and Cohen}{Liu and Cohen}{}]{liu_origin_2017}
Shi Liu and R.~E. Cohen.
\newblock
\newblock ``Origin of {Negative} {Longitudinal} {Piezoelectric} {Effect}.''  {\it Physical Review Letters\/}~119, no. 20 (2017): 207601.

\bibitem[\protect\citeauthoryear{Qi and Rappe}{Qi and Rappe}{}]{qi_widespread_2021}
Yubo Qi and Andrew~M. Rappe.
\newblock
\newblock ``Widespread {Negative} {Longitudinal} {Piezoelectric} {Responses} in {Ferroelectric} {Crystals} with {Layered} {Structures}.''  {\it Physical Review Letters\/}~126, no. 21 (2021): 217601.

\bibitem[\protect\citeauthoryear{You, Zhang, Zhou, Chaturvedi, Morris, Liu, Chang, Ichinose, Funakubo, Hu, Wu, Liu, Dong, and Wang}{You et~al.}{}]{you_origin_2019}
Lu~You, Yang Zhang, Shuang Zhou, Apoorva Chaturvedi, Samuel~A. Morris, Fucai Liu, Lei Chang, Daichi Ichinose, Hiroshi Funakubo, Weijin Hu, Tom Wu, Zheng Liu, Shuai Dong, and Junling Wang.
\newblock
\newblock ``Origin of giant negative piezoelectricity in a layered van der {Waals} ferroelectric.''  {\it Science Advances\/}~5, no. 4 (2019): eaav3780.

\bibitem[\protect\citeauthoryear{Yao, Bai, Jin, Zhang, Zheng, Xu, Chen, Wang, Liu, Wang, and Zhu}{Yao et~al.}{}]{yao_anomalous_2023}
Xiaodong Yao, Yinxin Bai, Cheng Jin, Xinyu Zhang, Qunfei Zheng, Zedong Xu, Lang Chen, Shanmin Wang, Ying Liu, Junling Wang, and Jinlong Zhu.
\newblock
\newblock ``Anomalous polarization enhancement in a van der {Waals} ferroelectric material under pressure.''  {\it Nature Communications\/}~14, no. 1 (2023): 4301.

\bibitem[\protect\citeauthoryear{Ding, Chen, Gui, You, Yao, and Dong}{Ding et~al.}{}]{ding_phase_2021}
Ning Ding, Jun Chen, Churen Gui, Haipeng You, Xiaoyan Yao, and Shuai Dong.
\newblock
\newblock ``Phase competition and negative piezoelectricity in interlayer-sliding ferroelectric {{ZrI$_2$}}.''  {\it Physical Review Materials\/}~5, no. 8 (2021): 084405.

\bibitem[\protect\citeauthoryear{Xu, Yang, Liu, Wang, and Wang}{Xu et~al.}{}]{xu_how_2024}
Jinrong Xu, Ziyue Yang, Wenjing Liu, Li~Wang, and Ying Wang.
\newblock
\newblock ``How to enhance the polarization intensity of two-dimensional sliding ferroelectricity for hexagonal boron- or nitrogen-based binary compounds?.''  {\it Journal of Physics: Condensed Matter\/}~36, no. 20 (2024): 205505.

\bibitem[\protect\citeauthoryear{Jaikissoon, K{\"o}ro{\u g}lu, Yang, Neilson, Saraswat, and Pop}{Jaikissoon et~al.}{}]{jaikissoon2024cmoscompatible}
Marc Jaikissoon, {\c C}a{\u g}{\i}l K{\"o}ro{\u g}lu, Jerry~A. Yang, Kathryn Neilson, Krishna~C. Saraswat, and Eric Pop.
\newblock
\newblock ``{{CMOS-compatible}} Strain Engineering for Monolayer Semiconductor Transistors.''  {\it Nature Electronics\/}~7, no. 10 (2024): 885--891.

\bibitem[\protect\citeauthoryear{Kim, Yankowitz, Fallahazad, Kang, Movva, Huang, Larentis, Corbet, Taniguchi, Watanabe, Banerjee, LeRoy, and Tutuc}{Kim et~al.}{}]{kim2016van}
Kyounghwan Kim, Matthew Yankowitz, Babak Fallahazad, Sangwoo Kang, Hema C.~P. Movva, Shengqiang Huang, Stefano Larentis, Chris~M. Corbet, Takashi Taniguchi, Kenji Watanabe, Sanjay~K. Banerjee, Brian~J. LeRoy, and Emanuel Tutuc.
\newblock
\newblock ``Van Der {{Waals Heterostructures}} with {{High Accuracy Rotational Alignment}}.''  {\it Nano Letters\/}~16, no. 3 (2016): 1989--1995.

\bibitem[\protect\citeauthoryear{Cao, Luo, Fatemi, Fang, {Sanchez-Yamagishi}, Watanabe, Taniguchi, Kaxiras, and {Jarillo-Herrero}}{Cao et~al.}{}]{cao2016superlatticeinduced}
Y.~Cao, J.~Y. Luo, V.~Fatemi, S.~Fang, J.~D. {Sanchez-Yamagishi}, K.~Watanabe, T.~Taniguchi, E.~Kaxiras, and P.~{Jarillo-Herrero}.
\newblock
\newblock ``Superlattice-{{Induced Insulating States}} and {{Valley-Protected Orbits}} in {{Twisted Bilayer Graphene}}.''  {\it Physical Review Letters\/}~117, no. 11 (2016): 116804.

\bibitem[\protect\citeauthoryear{Murata, Yoshino, Yadav, Honda, and Shirakawa}{Murata et~al.}{}]{murata1997pt}
Keizo Murata, Harukazu Yoshino, Hari~Om Yadav, Yoshiaki Honda, and Naoki Shirakawa.
\newblock
\newblock ``Pt Resistor Thermometry and Pressure Calibration in a Clamped Pressure Cell with the Medium, {{Daphne}} 7373.''  {\it Review of Scientific Instruments\/}~68, no. 6 (1997): 2490--2493.

\bibitem[\protect\citeauthoryear{Kamat, Dans, Saha, Kokovin, Paglione, Schmalian, and Ramshaw}{Kamat et~al.}{}]{kamatthermodynamic2026}
Sahas Kamat, Jared Dans, Shanta Saha, Artem~D. Kokovin, Johnpierre Paglione, J{\"o}rg Schmalian, and B.~J. Ramshaw.
\newblock ``Thermodynamic Discovery of Tetracriticality and Emergent Multicomponent Superconductivity in {UTe}$_2$.''  .
\newblock {\it arXiv preprint arXiv:2603.17905\/}.

\bibitem[\protect\citeauthoryear{Soler, Artacho, Gale, Garc{\'i}a, Junquera, Ordej{\'o}n, and {S{\'a}nchez-Portal}}{Soler et~al.}{}]{solerSIESTAMethodInitio2002}
Jos{\'e}~M. Soler, Emilio Artacho, Julian~D. Gale, Alberto Garc{\'i}a, Javier Junquera, Pablo Ordej{\'o}n, and Daniel {S{\'a}nchez-Portal}.
\newblock
\newblock ``The {{SIESTA}} Method for Ab Initio Order-{{N}} Materials Simulation.''  {\it Journal of Physics: Condensed Matter\/}~14, no. 11 (2002): 2745.

\bibitem[\protect\citeauthoryear{Garc{\'i}a, Verstraete, Pouillon, and Junquera}{Garc{\'i}a et~al.}{}]{garciaPsmlFormatLibrary2018}
Alberto Garc{\'i}a, Matthieu~J. Verstraete, Yann Pouillon, and Javier Junquera.
\newblock
\newblock ``The Psml Format and Library for Norm-Conserving Pseudopotential Data Curation and Interoperability.''  {\it Computer Physics Communications\/}227 (2018): 51--71.

\bibitem[\protect\citeauthoryear{van Setten, Giantomassi, Bousquet, Verstraete, Hamann, Gonze, and Rignanese}{van Setten et~al.}{}]{pseudodojo}
M.~J. van Setten, Matteo Giantomassi, Eric Bousquet, Matthieu~J. Verstraete, D.~R. Hamann, Xavier Gonze, and G.-M. Rignanese.
\newblock
\newblock ``The {PseudoDojo}: Training and Grading a 85 Element Optimized Norm-Conserving Pseudopotential Table.''  {\it Computer Physics Communications\/}226 (2018): 39--54.

\bibitem[\protect\citeauthoryear{Junquera, Paz, {S{\'a}nchez-Portal}, and Artacho}{Junquera et~al.}{}]{junqueraNumericalAtomicOrbitals2001}
Javier Junquera, {\'O}scar Paz, Daniel {S{\'a}nchez-Portal}, and Emilio Artacho.
\newblock
\newblock ``Numerical Atomic Orbitals for Linear-Scaling Calculations.''  {\it Physical Review B\/}~64, no. 23 (2001): 235111.

\bibitem[\protect\citeauthoryear{Bennett, Pizzochero, Junquera, and Kaxiras}{Bennett et~al.}{}]{bennettAccurateEfficientLocalized2025}
Daniel Bennett, Michele Pizzochero, Javier Junquera, and Efthimios Kaxiras.
\newblock
\newblock ``Accurate and Efficient Localized Basis Sets for Two-Dimensional Materials.''  {\it Physical Review B\/}~111, no. 12 (2025): 125123.

\bibitem[\protect\citeauthoryear{Perdew, Burke, and Ernzerhof}{Perdew et~al.}{}]{pbe}
John~P. Perdew, Kieron Burke, and Matthias Ernzerhof.
\newblock
\newblock ``Generalized Gradient Approximation Made Simple.''  {\it Phys. Rev. Lett.\/}77 (1996): 3865--3868.

\bibitem[\protect\citeauthoryear{Grimme, Antony, Ehrlich, and Krieg}{Grimme et~al.}{}]{grimme2010consistent}
Stefan Grimme, Jens Antony, Stephan Ehrlich, and Helge Krieg.
\newblock
\newblock ``A Consistent and Accurate \emph{ab initio} Parametrization of Density Functional Dispersion Correction ({DFT-D}) for the 94 Elements {H--Pu}.''  {\it The Journal of Chemical Physics\/}~132, no. 15 (2010): 154104.

\bibitem[\protect\citeauthoryear{Zhu, Bennett, Larson, Ezzi, Manousakis, and Kaxiras}{Zhu et~al.}{}]{zhuWavefunctionTexturesTwisted2026}
Albert Zhu, Daniel Bennett, Daniel~T. Larson, Mohammed M.~Al Ezzi, Efstratios Manousakis, and Efthimios Kaxiras.
\newblock
\newblock ``Wavefunction Textures in Twisted Bilayer Graphene from First Principles.''  {\it Physical Review B\/}~113, no. 4 (2026): L041112.

\bibitem[\protect\citeauthoryear{Rutter}{Rutter}{}]{rutterC2xToolVisualisation2018}
M.~J. Rutter.
\newblock
\newblock ``C2x: {{A}} Tool for Visualisation and Input Preparation for {{Castep}} and Other Electronic Structure Codes.''  {\it Computer Physics Communications\/}225 (2018): 174--179.

\end{thebibliography}

\bmsubsection*{Supporting Information}

Additional supporting information can be found online in the Supporting Information
section.

\begin{figure*}[t]
    \centering
    \includegraphics[width=\textwidth]{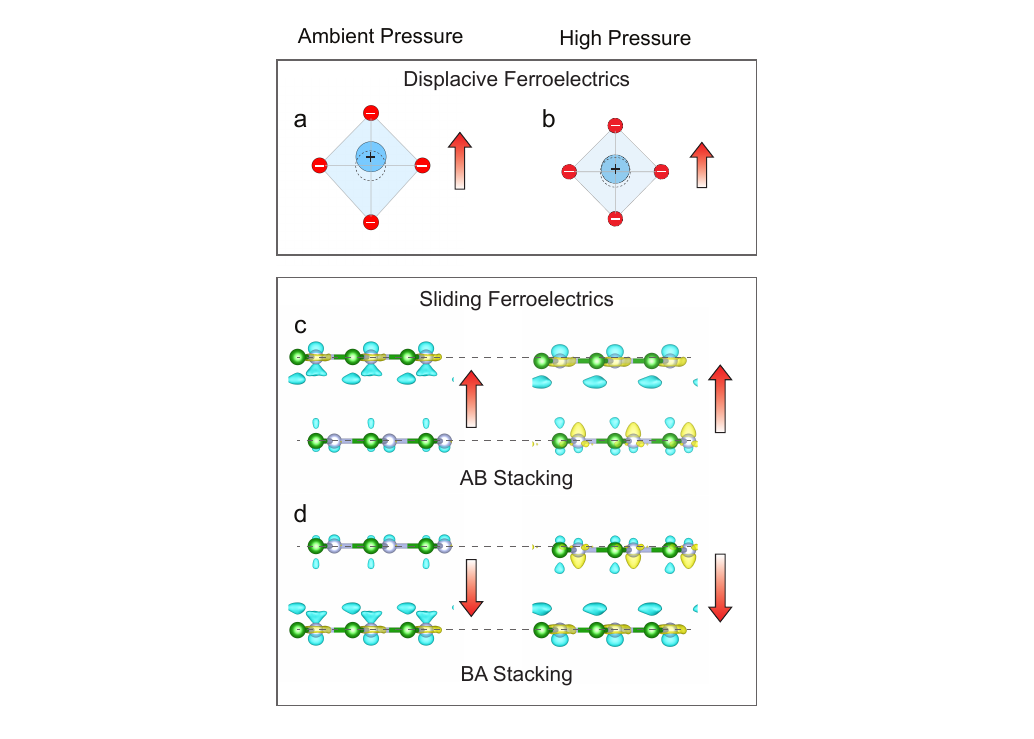}
    \caption{
    \textbf{Pressure responses of conventional and sliding ferroelectrics.}
    (a) Schematic of a conventional displacive ferroelectric at ambient pressure, where polarization originates from the off-center displacement of the positively charged ion relative to the surrounding negatively charged ions.
    (b) Under high pressure, the ionic displacement is reduced, and the polarization decreases.
    (c,d) Calculated structures and charge-density-difference distributions of bilayer rBN in the AB and BA registries, respectively.
    The left and right columns show the calculated results at 0 and 1.5~GPa, respectively.
    Yellow and cyan isosurfaces indicate positive and negative charge-density differences, corresponding to electron accumulation and depletion, respectively, at
    $\Delta\rho=\pm 0.002\,e\,\text{\AA}^{-3}$.
    The horizontal dashed lines mark the layer positions at 0~GPa and serve as guides to illustrate the pressure-induced reduction of the interlayer spacing.
    The arrows indicate the out-of-plane polarization directions, while the lengths are schematic.
    }
    \label{fig:fig1}
\end{figure*}

\begin{figure*}[t]
    \centering
    \includegraphics[width=\textwidth]{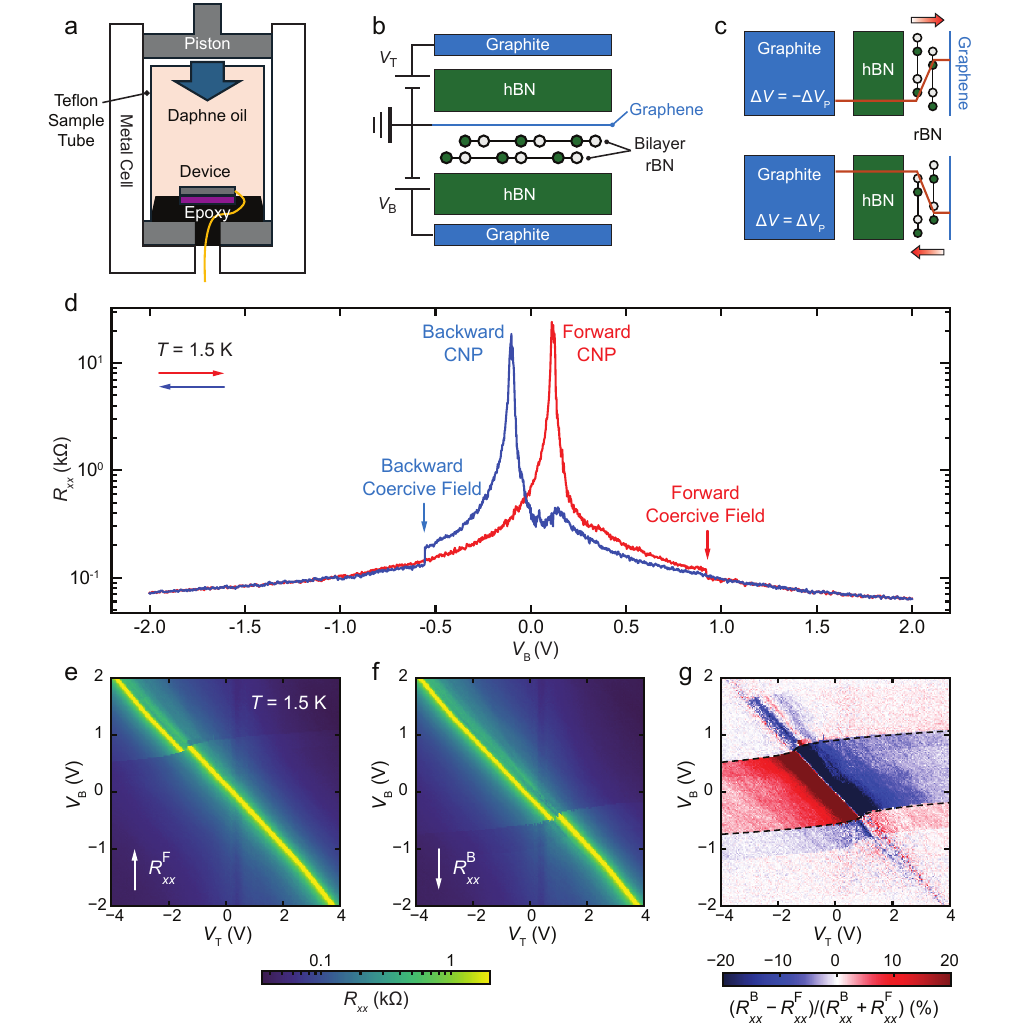}
    \caption{
    \textbf{Device geometry and ambient-pressure graphene charge sensing of sliding ferroelectricity.}
    (a) Schematic of the high-pressure measurement setup.
    (b) Device geometry of the graphene/bilayer rBN heterostructure.
    (c) Electrostatic potential profile illustrating the polarization-induced offset generated by two opposite stacking registries.
    (d) Forward and backward bottom-gate sweeps of graphene longitudinal resistance at ambient pressure. The forward and backward CNPs and coercive fields are labeled.
    (e,f) Dual-gate resistance maps measured during forward and backward bottom-gate sweeps.
    (g) Normalized hysteretic component of the resistance. The black dashed lines indicate the switching boundaries corresponding to a constant electric field across the bottom hBN and bilayer rBN.
    }
    \label{fig:fig2}
\end{figure*}

\begin{figure*}[t]
    \centering
    \includegraphics[width=\textwidth]{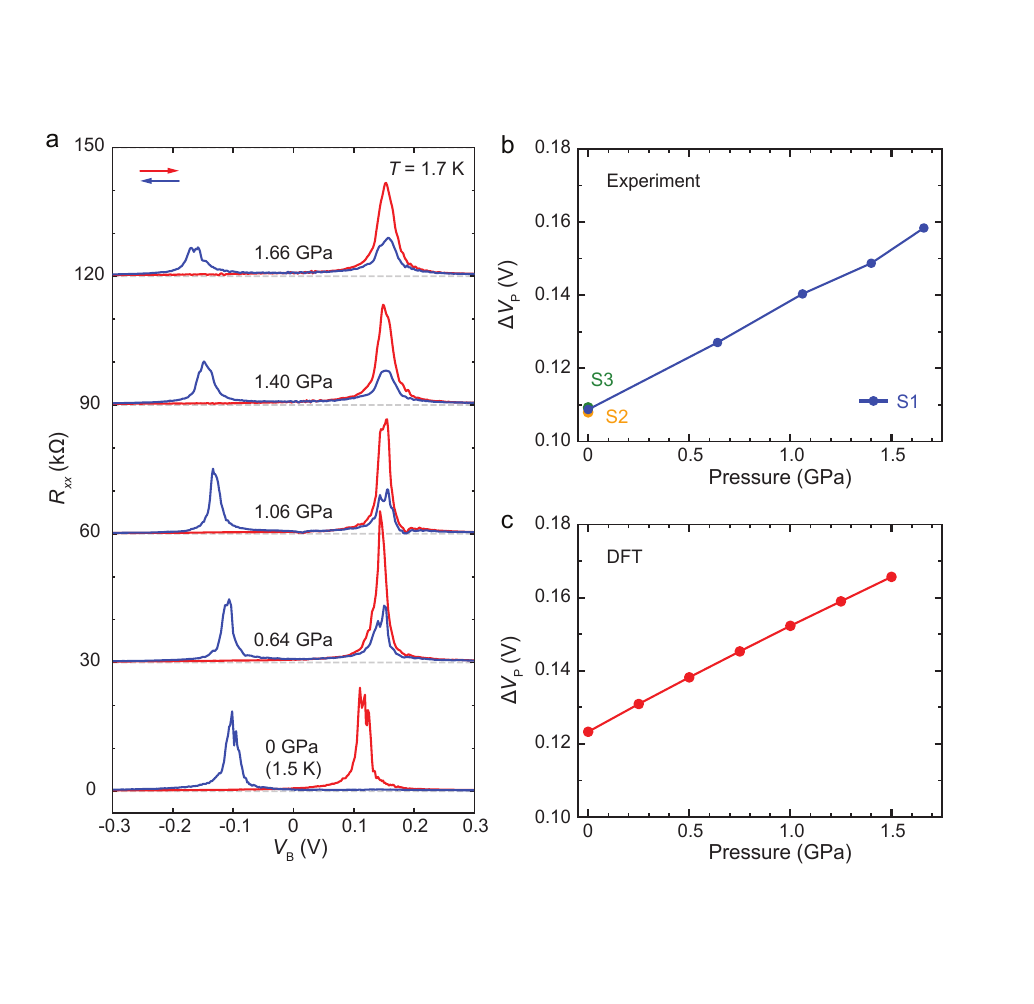}
    \caption{
    \textbf{Pressure-enhanced charge-neutrality-point separation.}
    (a) Forward and backward bottom-gate sweeps of graphene resistance at different hydrostatic pressures.
    The traces are vertically offset by \(30\,\mathrm{k}\Omega\) for clarity.
    (b) Extracted built-in interlayer potential, \(\Delta V_{\mathrm{P}}\), as a function of pressure. The ambient-pressure values obtained from S2 and S3 are also included for comparison.
    (c) Calculated pressure dependence of $\Delta V_{\mathrm{P}}$ from first-principles calculations.
    }
    \label{fig:fig3}
\end{figure*}

\begin{figure*}[t]
    \centering
    \includegraphics[width=\textwidth]{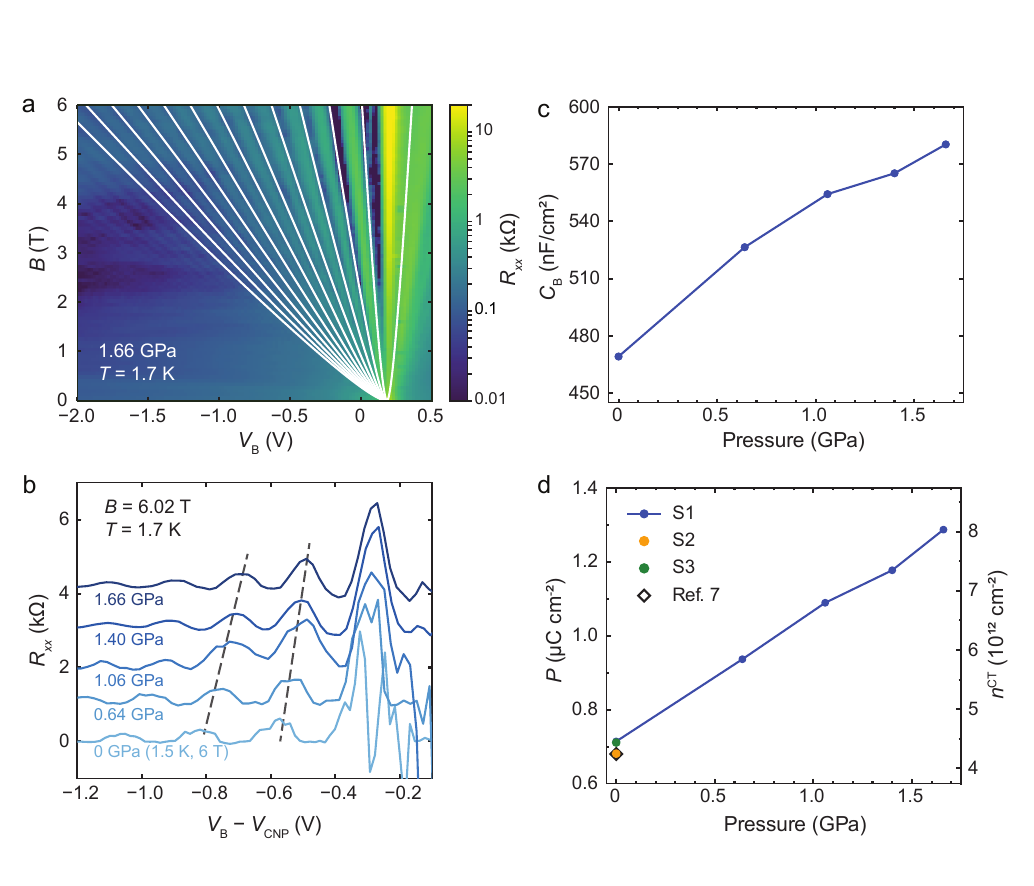}
    \caption{
    \textbf{Pressure-dependent bottom-gate capacitance and interlayer charge transfer.}
    (a) Graphene Landau fan measured at $1.66~\mathrm{GPa}$ and $1.7~\mathrm{K}$. The white lines are fits to the Landau-fan features at different filling factors, accounting for the quantum capacitance of graphene.
    (b) Line cuts at $B=6.02~\mathrm{T}$ for different pressures, showing the pressure-induced shift of the Shubnikov-de Haas peaks, as indicated by the gray dashed guide lines. The $0~\mathrm{GPa}$ data were measured at $B=6~\mathrm{T}$.
    The traces are vertically offset for clarity.
    (c) Extracted bottom-gate capacitance, $C_\mathrm{B}$, as a function of pressure.
    (d) Pressure dependence of the polarization \(P\) and interlayer charge-transfer density \(n_{\mathrm{CT}}\). Ambient-pressure values obtained from S2 and S3, together with the value reported in Ref.~7, are included for comparison.
    }
    \label{fig:fig4}
\end{figure*}


\end{document}